\documentclass[times,twocolumn]{aastex62}

\usepackage{cases}
\usepackage{graphicx}
\usepackage{subfigure}
\usepackage{natbib}
\usepackage{color}
\usepackage{tabularx}
\usepackage{bm}
\usepackage{threeparttable}
\newcommand{\thetae}{\theta_{\rm E}}
\newcommand{\pie}{\pi_{\rm E}}

\newcommand{\te}{t_{\rm E}}
\newcommand{\one}{OGLE-2017-BLG-1186}
\newcommand{\two}{OGLE-2017-BLG-1254}
\newcommand{\three}{OGLE-2017-BLG-1161}

\newcommand{\Sp}{{\it Spitzer}}
\DeclareGraphicsExtensions{.pdf,.png,.jpg}

\shorttitle{}
\shortauthors{Zang et al.}

\begin{document}

\title{{\large \Sp\ Microlensing parallax reveals two isolated stars in the Galactic bulge}}

\correspondingauthor{Weicheng Zang}
\email{zangwc17@mails.tsinghua.edu.cn}

\author[0000-0001-6000-3463]{Weicheng Zang}
\affiliation{Physics Department and Tsinghua Centre for Astrophysics, Tsinghua University, Beijing 100084, China}

\author{Yossi Shvartzvald}
\affiliation{IPAC, Mail Code 100-22, Caltech, 1200 E. California Blvd., Pasadena, CA 91125, USA}

\author{Tianshu Wang}
\affiliation{Physics Department and Tsinghua Centre for Astrophysics, Tsinghua University, Beijing 100084, China}

\author{Andrzej Udalski}
\affiliation{Warsaw University Observatory, Al. Ujazdowskie 4, 00-478 Warszawa, Poland}

\author{Chung-Uk Lee}
\affiliation{Korea Astronomy and Space Science Institute, Daejon 34055, Republic of Korea}
\affiliation{University of Science and Technology, Korea, (UST), 217 Gajeong-ro Yuseong-gu, Daejeon 34113, Republic of Korea}

\author{Takahiro Sumi}
\affiliation{Department of Earth and Space Science, Graduate School of Science, Osaka University, Toyonaka, Osaka 560-0043, Japan}

\author{Jesper Skottfelt}
\affiliation{Centre for Electronic Imaging, Department of Physical Sciences, The Open University, Milton Keynes, MK7 6AA, UK}

\author[0000-0001-9952-7408]{Shun-Sheng Li}
\affiliation{National Astronomical Observatories, Chinese Academy of Sciences, Beijing 100101, China}
\affiliation{School of Astronomy and Space Science, University of Chinese Academy of Sciences, Beijing 100049, China}

\author{Shude Mao}
\affiliation{Physics Department and Tsinghua Centre for Astrophysics, Tsinghua University, Beijing 100084, China}
\affiliation{National Astronomical Observatories, Chinese Academy of Sciences, Beijing 100101, China}

\author{Wei Zhu}
\affiliation{Canadian Institute for Theoretical Astrophysics, University of Toronto, 60 St George Street, Toronto, ON M5S 3H8, Canada}

\nocollaboration

% Spitzer
\author{Jennifer~C.~Yee}
\affiliation{Center for Astrophysics | Harvard \& Smithsonian, 60 Garden St.,Cambridge, MA 02138, USA}

\author{Sebastiano Calchi Novati}
\affiliation{IPAC, Mail Code 100-22, Caltech, 1200 E. California Blvd., Pasadena, CA 91125, USA}

\author{Charles A. Beichman}
\affiliation{IPAC, Mail Code 100-22, Caltech, 1200 E. California Blvd., Pasadena, CA 91125, USA}

\author{Geoffery Bryden}
\affiliation{Jet Propulsion Laboratory, California Institute of Technology, 4800 Oak Grove Drive, Pasadena, CA 91109, USA}

\author{Sean Carey}
\affiliation{IPAC, Mail Code 100-22, Caltech, 1200 E. California Blvd., Pasadena, CA 91125, USA}

\author{B.~Scott~Gaudi}
\affiliation{Department of Astronomy, Ohio State University, 140 W. 18th Ave., Columbus, OH  43210, USA}

\author{Calen~B.~Henderson}
\affiliation{IPAC, Mail Code 100-22, Caltech, 1200 E. California Blvd., Pasadena, CA 91125, USA}

\collaboration{(The \emph{Spitzer} Team)}

% ogle

\author{Przemek Mr\'{o}z}
\affiliation{Warsaw University Observatory, Al. Ujazdowskie 4, 00-478 Warszawa, Poland}

\author{Jan~Skowron}
\affiliation{Warsaw University Observatory, Al. Ujazdowskie 4, 00-478 Warszawa, Poland}

\author{Radoslaw~Poleski}
\affiliation{Warsaw University Observatory, Al. Ujazdowskie 4, 00-478 Warszawa, Poland}
\affiliation{Department of Astronomy, Ohio State University, 140 W. 18th Ave., Columbus, OH  43210, USA}

\author{Micha{\l}~K.~Szyma\'{n}ski}
\affiliation{Warsaw University Observatory, Al. Ujazdowskie 4, 00-478 Warszawa, Poland}

\author{Igor Soszy\'{n}ski}
\affiliation{Warsaw University Observatory, Al. Ujazdowskie 4, 00-478 Warszawa, Poland}

\author{Pawe{\l} Pietrukowicz}
\affiliation{Warsaw University Observatory, Al. Ujazdowskie 4, 00-478 Warszawa, Poland}

\author{Szymon Koz{\l}owski}
\affiliation{Warsaw University Observatory, Al. Ujazdowskie 4, 00-478 Warszawa, Poland}

\author{Krzysztof Ulaczyk}
\affiliation{Department of Physics, University of Warwick, Gibbet Hill Road, Coventry, CV4~7AL,~UK}

\author{Krzysztof A.~Rybicki}
\affiliation{Warsaw University Observatory, Al. Ujazdowskie 4, 00-478 Warszawa, Poland}

\author{Patryk Iwanek}
\affiliation{Warsaw University Observatory, Al. Ujazdowskie 4, 00-478 Warszawa, Poland}
\collaboration{(The OGLE Collaboration)}

\author{Etienne Bachelet}
\affiliation{Las Cumbres Observatory, 6740 Cortona Drive, suite 102, Goleta, CA 93117, USA}

\author{Grant Christie}
\affiliation{Auckland Observatory, Auckland, New Zealand}

\author{Jonathan Green}
\affiliation{Kumeu Observatory, Kumeu, New Zealand}

\author{Steve Hennerley}
\affiliation{Kumeu Observatory, Kumeu, New Zealand}

\author{Dan Maoz}
\affiliation{School of Physics and Astronomy, Tel-Aviv University, Tel-Aviv 6997801, Israel}

\author{Tim Natusch}
\affiliation{Auckland Observatory, Auckland, New Zealand}
\affiliation{Institute for Radio Astronomy and Space Research (IRASR), AUT University, Auckland, New Zealand}

\author{Richard W. Pogge}
\affiliation{Department of Astronomy, Ohio State University, 140 W. 18th Ave., Columbus, OH  43210, USA}
\affiliation{Center for Cosmology \& AstroParticle Physics, The Ohio State University, 191 West Woodruff Avenue, Columbus, OH 43210}
\author{Rachel  A. Street}
\affiliation{Las Cumbres Observatory, 6740 Cortona Drive, suite 102, Goleta, CA 93117, USA}

\author{Yiannis Tsapras}
\affiliation{Astronomisches Rechen-Institut, Zentrum f{\"u}r Astronomie der Universit{\"a}t Heidelberg (ZAH), 69120 Heidelberg, Germany}

\collaboration{(The LCO and $\mu$FUN Follow-up Teams)}

% kmt
\author{Michael D. Albrow}
\affiliation{University of Canterbury, Department of Physics and Astronomy, Private Bag 4800, Christchurch 8020, New Zealand}

\author{Sun-Ju Chung}
\affiliation{Korea Astronomy and Space Science Institute, Daejon 34055, Republic of Korea}
\affiliation{Korea University of Science and Technology, 217 Gajeong-ro, Yuseong-gu, Daejeon 34113, Republic of Korea}

\author{Andrew Gould}
\affiliation{Korea Astronomy and Space Science Institute, Daejon 34055, Republic of Korea}
\affiliation{Department of Astronomy, Ohio State University, 140 W. 18th Ave., Columbus, OH 43210, USA}
\affiliation{Max-Planck-Institute for Astronomy, K\"onigstuhl 17, 69117 Heidelberg, Germany}

\author{Cheongho Han}
\affiliation{Department of Physics, Chungbuk National University, Cheongju 28644, Republic of Korea}

\author{Kyu-Ha Hwang}
\affiliation{Korea Astronomy and Space Science Institute, Daejon 34055, Republic of Korea}

\author{Youn Kil Jung}
\affiliation{Harvard-Smithsonian Center for Astrophysics, 60 Garden St.,Cambridge, MA 02138, USA}
\affiliation{Korea Astronomy and Space Science Institute, Daejon 34055, Republic of Korea}

\author{Yoon-Hyun Ryu}
\affiliation{Korea Astronomy and Space Science Institute, Daejon 34055, Republic of Korea}

\author{In-Gu Shin}
\affiliation{Korea Astronomy and Space Science Institute, Daejon 34055, Republic of Korea}

\author{Sang-Mok Cha}
\affiliation{Korea Astronomy and Space Science Institute, Daejon 34055, Republic of Korea}
\affiliation{School of Space Research, Kyung Hee University, Yongin, Kyeonggi 17104, Republic of Korea} 

\author{Dong-Jin Kim}
\affiliation{Korea Astronomy and Space Science Institute, Daejon 34055, Republic of Korea}

\author{Hyoun-Woo Kim}
\affiliation{Korea Astronomy and Space Science Institute, Daejon 34055, Republic of Korea}

\author{Seung-Lee Kim}
\affiliation{Korea Astronomy and Space Science Institute, Daejon 34055, Republic of Korea}
\affiliation{Korea University of Science and Technology, 217 Gajeong-ro, Yuseong-gu, Daejeon 34113, Republic of Korea}

\author{Dong-Joo Lee}
\affiliation{Korea Astronomy and Space Science Institute, Daejon 34055, Republic of Korea}

\author{Yongseok Lee}
\affiliation{Korea Astronomy and Space Science Institute, Daejon 34055, Republic of Korea}
\affiliation{School of Space Research, Kyung Hee University, Yongin, Kyeonggi 17104, Republic of Korea}

\author{Byeong-Gon Park}
\affiliation{Korea Astronomy and Space Science Institute, Daejon 34055, Republic of Korea}
\affiliation{Korea University of Science and Technology, 217 Gajeong-ro, Yuseong-gu, Daejeon 34113, Republic of Korea}

\author{Richard W. Pogge}
\affiliation{Department of Astronomy, Ohio State University, 140 W. 18th Ave., Columbus, OH 43210, USA}

\collaboration{(The KMTNet Collaboration)}

% moa
\author{Ian~A.~Bond}
\affiliation{Institute of Natural and Mathematical Sciences, Massey University, Auckland 0745, New Zealand}

\author{Fumio~Abe}
\affiliation{Institute for Space-Earth Environmental Research, Nagoya University, Nagoya 464-8601, Japan}

\author{Richard Barry}
\affiliation{Code 667, NASA Goddard Space Flight Center, Greenbelt, MD 20771, USA}

\author{David~P.~Bennett}
\affiliation{Code 667, NASA Goddard Space Flight Center, Greenbelt, MD 20771, USA}
\affiliation{Department of Astronomy, University of Maryland, College Park, MD 20742, USA}

\author{Aparna~Bhattacharya}
\affiliation{Code 667, NASA Goddard Space Flight Center, Greenbelt, MD 20771, USA}
\affiliation{Department of Astronomy, University of Maryland, College Park, MD 20742, USA}

\author{Martin~Donachie}
\affiliation{Department of Physics, University of Auckland, Private Bag 92019, Auckland, New Zealand}

\author{Akihiko~Fukui}
\affiliation{Okayama Astrophysical Observatory, National Astronomical Observatory of Japan, 3037-5 Honjo, Kamogata, Asakuchi, Okayama 719-0232, Japan}

\author{Yuki~Hirao}
\affiliation{Department of Earth and Space Science, Graduate School of Science, Osaka University, Toyonaka, Osaka 560-0043, Japan}

\author{Yoshitaka~Itow}
\affiliation{Institute for Space-Earth Environmental Research, Nagoya University, Nagoya 464-8601, Japan}

\author{Iona~Kondo}
\affiliation{Department of Earth and Space Science, Graduate School of Science, Osaka University, Toyonaka, Osaka 560-0043, Japan}

\author{Naoki~Koshimoto}
\affiliation{Department of Astronomy, Graduate School of Science, The University of Tokyo, 7-3-1 Hongo, Bunkyo-ku, Tokyo 113-0033, Japan}
\affiliation{National Astronomical Observatory of Japan, 2-21-1 Osawa, Mitaka, Tokyo 181-8588, Japan}

\author{Man~Cheung~Alex~Li}
\affiliation{Department of Physics, University of Auckland, Private Bag 92019, Auckland, New Zealand}

\author{Yutaka~Matsubara}
\affiliation{Institute for Space-Earth Environmental Research, Nagoya University, Nagoya 464-8601, Japan}

\author{Yasushi~Muraki}
\affiliation{Institute for Space-Earth Environmental Research, Nagoya University, Nagoya 464-8601, Japan}

\author{Shota~Miyazaki}
\affiliation{Department of Earth and Space Science, Graduate School of Science, Osaka University, Toyonaka, Osaka 560-0043, Japan}

\author{Masayuki~Nagakane}
\affiliation{Department of Earth and Space Science, Graduate School of Science, Osaka University, Toyonaka, Osaka 560-0043, Japan}

\author{Cl\'ement~Ranc}
\affiliation{Code 667, NASA Goddard Space Flight Center, Greenbelt, MD 20771, USA}

\author{Nicholas~J.~Rattenbury}
\affiliation{Department of Physics, University of Auckland, Private Bag 92019, Auckland, New Zealand}

\author{Haruno~Suematsu}
\affiliation{Department of Earth and Space Science, Graduate School of Science, Osaka University, Toyonaka, Osaka 560-0043, Japan}

\author{Denis~J.~Sullivan}
\affiliation{School of Chemical and Physical Sciences, Victoria University, Wellington, New Zealand}

\author{Daisuke~Suzuki}
\affiliation{Institute of Space and Astronautical Science, Japan Aerospace Exploration Agency, 3-1-1 Yoshinodai, Chuo, Sagamihara, Kanagawa, 252-5210, Japan}

\author{Paul~J.~Tristram}
\affiliation{University of Canterbury Mt.\ John Observatory, P.O. Box 56, Lake Tekapo 8770, New Zealand}

\author{Atsunori~Yonehara}
\affiliation{Department of Physics, Faculty of Science, Kyoto Sangyo University, 603-8555 Kyoto, Japan}
\collaboration{(The MOA Collaboration)}

% mindstep
\author{Martin Dominik}
\affiliation{Centre for Exoplanet Science, SUPA, School of Physics \& Astronomy, University of St Andrews, North Haugh, St Andrews KY16 9SS, UK}

\author{Markus Hundertmark}
\affiliation{Astronomisches Rechen-Institut, Zentrum f{\"u}r Astronomie der Universit{\"a}t Heidelberg (ZAH), 69120 Heidelberg, Germany}

\author{Uffe G. J{\o}rgensen}
\affiliation{Niels Bohr Institute \& Centre for Star and Planet Formation, University of Copenhagen, {\O}ster Voldgade 5, 1350 Copenhagen, Denmark}

\author{Sohrab Rahvar}
\affiliation{Department of Physics, Sharif University of Technology, PO Box 11155-9161 Tehran, Iran}

\author{Sedighe Sajadian}
\affiliation{Department of Physics, Isfahan University of Technology, Isfahan 84156-83111, Iran}

\author{Colin Snodgrass}
\affiliation{School of Physical Sciences, Faculty of Science, Technology, Engineering and Mathematics, The Open University, Walton Hall, Milton Keynes, MK7 6AA, UK} 

\author{Valerio~Bozza}
\affiliation{Dipartimento di Fisica "E.R. Caianiello", Universit{\`a} di Salerno, Via Giovanni Paolo II 132, 84084, Fisciano, Italy}
\affiliation{Istituto Nazionale di Fisica Nucleare, Sezione di Napoli, Napoli, Italy}

\author{Martin J. Burgdorf} 
\affiliation{Universit{\"a}t Hamburg, Faculty of Mathematics, Informatics and Natural Sciences, Department of Earth Sciences, Meteorological Institute, Bundesstra\ss{}e 55, 20146 Hamburg, Germany}

\author{Daniel F. Evans}
\affiliation{Astrophysics Group, Keele University, Staffordshire, ST5 5BG, UK}

\author{Roberto Figuera Jaimes}
\affiliation{Astronomisches Rechen-Institut, Zentrum f{\"u}r Astronomie der Universit{\"a}t Heidelberg (ZAH), 69120 Heidelberg, Germany}

\author{Yuri I. Fujii} 
\affiliation{Niels Bohr Institute \& Centre for Star and Planet Formation, University of Copenhagen, {\O}ster Voldgade 5, 1350 Copenhagen, Denmark}
\affiliation{Institute for Advanced Research, Nagoya University, Furo-cho, Chikusa-ku, Nagoya, 464-8601, Japan}

\author{Luigi Mancini}
\affiliation{Max Planck Institute for Astronomy, K{\"o}nigstuhl 17, 69117 Heidelberg, Germany}
\affiliation{Dipartimento di Fisica, Università di Roma Tor Vergata, Via della Ricerca Scientifica 1, I-00133—Roma, Italy}
\affiliation{International Institute for Advanced Scientific Studies (IIASS), Via G. Pellegrino 19, I-84019, Vietri sul Mare (SA), Italy}

\author{Penelope Longa-Pe{\~n}a}
\affiliation{Centro de Astronom{\'{\i}}a (CITEVA), Universidad de Antofagasta, Avda.\ U.\ de Antofagasta 02800, Antofagasta, Chile}

\author{Christiane Helling}
\affiliation{Centre for Exoplanet Science, SUPA, School of Physics \& Astronomy, University of St Andrews, North Haugh, St Andrews KY16 9SS, UK}

\author{Nuno Peixinho}
\affiliation{CITEUC -- Center for Earth and Space Research of the University of Coimbra, Geophysical and Astronomical Observatory, R. Observat{\'{o}}rio s/n, 3040-004 Coimbra, Portugal}

\author{Markus Rabus} 
\affiliation{Las Cumbres Observatory Global Telescope, 6740 Cortona Dr., Suite 102, Goleta, CA 93111, USA} 
\affiliation{Department of Physics, University of California, Santa Barbara, CA 93106-9530, USA}

\author{John Southworth}
\affiliation{Astrophysics Group, Keele University, Staffordshire, ST5 5BG, UK}

\author{Eduardo Unda-Sanzana}
\affiliation{Centro de Astronom{\'{\i}}a (CITEVA), Universidad de Antofagasta, Avda.\ U.\ de Antofagasta 02800, Antofagasta, Chile}

\author{Carolina von Essen}
\affiliation{Stellar Astrophysics Centre, Department of Physics and Astronomy, Aarhus University, Ny Munkegade 120, 8000 Aarhus C, Denmark}
\collaboration{(The MiNDSTEp Collaboration)}

\begin{abstract}
We report the mass and distance measurements of two single-lens events from the 2017 \Sp\ microlensing campaign. The ground-based observations yield the detection of finite-source effects, and the microlens parallaxes are derived from the joint analysis of ground-based observations and \Sp\ observations. We find that the lens of \two\ is a $0.60 \pm 0.03 M_{\odot}$ star with $D_{\rm LS} = 0.53 \pm 0.11~\text{kpc}$, where $D_{\rm LS}$ is the distance between the lens and the source. The second event, \three, is subject to the known satellite parallax degeneracy, and thus is either a $0.51^{+0.12}_{-0.10} M_{\odot}$ star with $D_{\rm LS} = 0.40 \pm 0.12~\text{kpc}$ or a $0.38^{+0.13}_{-0.12} M_{\odot}$ star with $D_{\rm LS} = 0.53 \pm 0.19~\text{kpc}$. Both of the lenses are therefore isolated stars in the Galactic bulge. By comparing the mass and distance distributions of the eight published \Sp\ finite-source events with the expectations from a Galactic model, we find that the \Sp\ sample is in agreement with the probability of finite-source effects occurrence in single lens events.

\end{abstract}

\section{Introduction}\label{intro}

Gravitational microlensing opens a powerful window for probing isolated objects with various masses such as free-floating planets, brown dwarfs, low-mass stars and black holes. At the low-mass end, microlensing has detected several free-floating planet candidates \citep{Sumi2011, Mroz2017a, MrozNeptune,Mroz2FFP}, including a few possible Earth-mass objects. Such discoveries are crucial for testing theories about the origin and evolution of free-floating planets \citep{MaFFP, Clanton2017, Veras2012, Pfyffer2015, Barclay2017}. For more massive objects (i.e., isolated brown dwarfs), five have been discovered by microlensing: OGLE-2007-BLG-224L \citep{OB07224}, OGLE-2015-BLG-1268L \citep{ZhuPLFS}, OGLE-2015-BLG-1482\footnote{OGLE-2015-BLG-1482 has two possible solutions, with $M=55\pm9M_{J}$ or $M=96\pm23M_{J}$.} \citep{OB151482}, OGLE-2017-BLG-0896 \citep{OB170896}, and OGLE-2017-BLG-1186 \footnote{OGLE-2017-BLG-1186 has two possible solutions, with $M=45\pm1M_{J}$ or $M=73\pm2M_{J}$.} \citep{OB171186}. \cite{OB170896} recently announced the discovery of an isolated, extremely low-mass brown dwarf of $M\sim19M_{J}$, with proper motion in the opposite direction of disk stars, which indicates that it might be a halo brown dwarf or from a different, unknown counter-rotating population. At the high-mass end, \cite{Gouldremnant} estimated that $\sim20\%$ of microlensing events observed toward the Galactic bulge are caused by stellar remnants, and specifically that $\sim1\%$ are due to stellar-mass black holes, with another $\sim3\%$ due to neutron star lenses. The first observed example of this was the long-timescale ($\sim640$~day) event OGLE-1999-BUL-32, for which the microlens parallax measurement indicated this event could be a stellar black hole \citep{ShudeBH}. In addition, \cite{OGLE3BH} identified 13 microlensing events that are consistent with having a white-dwarf, neutron-star or a black-hole lens in the OGLE-III data base.  

%\sout{which provides a different sample for ``brown dwarf desert" \citep{Marcy2000} from other methods (transit \citep{Deleuil2008}, radial velocity \citep{Sahlman2011}, and direct imaging \citealt{Lafreniere2007}).}

In general, for microlensing events due to isolated lenses, the only measured parameter that describes the physical properties of the lens system is the Einstein timescale $\te$. Because $\te$ depends on the lens mass, the distances to the lens and source, and the transverse velocity (See Equation 17 of \citealt{Mao2012}), it can only be used to make a statistical estimate of the lens mass. Unambiguous measurements of the lens mass requires two second-order microlensing observables: the angular Einstein radius $\thetae$ and the microlens parallax $\pie$. For a lensing object, the total mass is related to the two observables by \citep{Gould1992, Gould2000}
\begin{equation}
    M_{\rm L} = \frac{\thetae}{{\kappa}\pie},
    \label{eq:mass}
\end{equation}
and its distance by
\begin{equation}
    D_{\rm L} = \frac{\mathrm{au}}{\pi_{\rm rel} + \pi_{\rm S}},\qquad \pi_{\rm rel} = \pie\thetae
\end{equation}
where $\kappa \equiv 4G/(c^2\mathrm{AU}) = 8.144$ mas$/M_{\odot}$, $\pi_{\rm S} = \mathrm{au}/D_{\rm S}$ is the source parallax, $D_{\rm S}$ is the source distance \citep{Gould1992,Gouldpies2004} and $\pi_{\rm rel}$ is the lens-source relative parallax. 

There are three methods to measure the microlens parallax $\pie$. The first one is ``orbital microlens parallax'', which can be measured when including the orbital motion of Earth around the Sun in modeling \citep{Gould1992, Alcock1995}. However, this method is generally feasible only for events with long microlensing timescales $\te\gtrsim$ year/$2\pi$ \citep[e.g.,][]{OB171434}. The second method, ``terrestrial microlens parallax'', in rare cases can be measured by a combination of simultaneous observations from ground-based telescopes that are well separated \citep[e.g.,][]{OB07224,OB080279}. The most efficient and robust method to measure the microlens parallax is to simultaneously observe an event from Earth and a satellite \citep{1966MNRAS.134..315R,1994ApJ...421L..75G}. That is the ``satellite microlens parallax''. The feasibility of satellite microlens parallax measurements has been demonstrated by \Sp\ microlensing programs \citep{2007ApJ...664..862D,2015ApJ...799..237U,2015ApJ...802...76Y,2015ApJ...805....8Z,Novati2015}. Since 2014, the \Sp\ satellite has observed more than 700 microlensing events toward the Galactic bulge, yielding the mass measurements of eight isolated lens objects \citep{ZhuPLFS, OB151482, OB161045, OB170896, OB171186}, including two in this work. 

For the measurements of the angular Einstein radius $\thetae$, \cite{Jan2017} recently reported the angular Einstein radius $\thetae$ measurement of microlensing event TCP J05074264+2447555 by interferometric resolution of the microlensed images. However, this method requires a rare, bright microlensing event (for TCP J05074264+2447555, $K \sim 10.6$ mag at the time of observation). Measurements of the angular Einstein radius $\thetae$ are obtained primarily via finite-source effects and an estimate of the angular diameter $\theta_{*}$ of the source from its de-reddened color and magnitude \citep[e.g.,][]{Kervella2008,Boyajian2014}
\begin{equation}\label{thetae} 
   \thetae = \frac{\theta_{*}}{\rho},
\end{equation}
where $\rho$ is the source size normalized by the Einstein radius, which can be measured from the modulation in the lensing light curve with finite-source effects. Such effects arise when the source transits a caustic (where the magnification diverges to infinity) or comes close to a cusp \citep{1994ApJ...421L..75G,Shude1994,Nemiroff1994}. Then the source cannot be regarded as a point-like source, and the observed magnification is the integration of the magnification pattern over the face of the source. Finite-source effects are frequently measured in binary/planetary events, for which the caustic structures are relatively large, but they are rarely measured in the case of a single lens event because the caustic is a single geometric point.

Here we present the mass and distance measurements of two \Sp\ single-lens microlensing events \three\ and \two. The ground-based observations yield a robust detection of finite-source effects for the two events, and the microlens parallaxes are derived from the joint analysis of ground-based observations and \Sp\ observations. Combining the measurements of $\thetae$ and $\pie$, we find that the lenses of the two events are both isolated stars in the Galactic bulge. The paper is structured as follows. In Section \ref{obser}, we introduce ground-based and \Sp\ observations of the two events. We then describe the light curve modeling process in Section \ref{model}, and present the physical parameters of the two events in Section \ref{lens}. Finally, our conclusions and the implications of our work are given in Section \ref{dis}.

%The probability of finite-source effect occurrence in a single-lens event is 
%\begin{equation}
    %P = \rho \equiv \frac{\theta_*}{\thetae} \propto \theta_*, 
%\end{equation}
%so the finite-source effect is strongly biased toward the sources that have large angular radius. 
% and the sources are both giant stars, with angular radius $\theta_* = ( , )$.

\section{Observations and Data reductions}\label{obser}

The observations of \three\ and \two\ both consist of \Sp, ground-based survey and ground-based follow-up observations.

%The \Sp\ observations in 2017 were carried out under a $\color{red}{?}$-hour program toward the Galactic bulge, whose scientific goal was to measure the Galactic distribution of planets at different stellar environments \citep{Novati2015,Zhu2017spitzer}.

The \Sp\ observations were part of a large program to measure the Galactic distribution of planets in different stellar environments \citep{Novati2015,Zhu2017spitzer}. The detailed protocols and strategies for the \Sp\ observations are discussed in \cite{YeeSpitzer}. Specifically, the two events were observed by the \Sp\ satellite because they were both high-magnification events, which are more sensitive to planets \citep{Griest1998}. The \Sp\ observations were taken using the 3.6 $\mu$m channel ($L-$band) of the IRAC camera.

%In brief, the strategies aims to select events if they (1) had or were likely to have significant sensitivity to planets, which is based on a combined consideration of the magnification \citep{Griest1998}, brightness, ground-based observation cadence and other features of the events; (2) could probably yield a parallax measurement. Events that meet the specified objective criteria were selected ``objectively'' and {\it must} be observed with a pre-specified cadence. Events that do not meet the criteria can still be chosen ``subjectively'' {\it at any time} for any reason,  but only data taken (or rather, made public) after this selection date can be used to calculate the planetary sensitivity of the events. In addition, events can be selected ``secretly" without any announcement and become ``subjectively" after the \Sp\ team makes a public announcement. Targets were submitted on Monday for observations on roughly Thursday through Wednesday for each of the six weeks of the \Sp\ campaign.

Ground-based surveys included the Optical Gravitational Lensing Experiment (OGLE, \citealt{OGLEIV}), the Microlensing Observations in Astrophysics (MOA, \citealt{MOA2016}), and the Korea Microlensing Telescope Network (KMTNet, \citealt{KMT2016}). OGLE is in its fourth phase (OGLE-IV), and the observations are carried out using its 1.3 m Warsaw Telescope equipped with a 1.4 ${\rm deg}^2$ FOV mosaic CCD camera at the Las Campanas Observatory in Chile.  The MOA group conducts a high cadence survey toward the Galactic bulge using its 1.8 m telescope equipped with a 2.2 ${\rm deg}^2$ FOV camera at the Mt. John University Observatory in New Zealand. KMTNet consists of three 1.6~m telescopes, equipped with 4 ${\rm deg}^2$ FOV cameras at the Cerro Tololo International Observatory (CTIO) in Chile (KMTC), the South African Astronomical Observatory (SAAO) in South Africa (KMTS), and the Siding Spring Observatory (SSO) in Australia (KMTA). The majority of observations were taken in the $I$-band for the OGLE and KMTNet groups, and the MOA-Red filter (which is similar to the sum of the standard Cousins $R$- and $I$-band filters) for the MOA group, with occasional observations taken in the $V$-band. 

The aim of the ground-based follow-up observations was to detect and characterize any planetary signatures with dense observations, which are crucial if an event is not heavily monitored by the ground-based surveys (e.g., \three) or the ground-based surveys could not observe due to weather \citep[e.g., OGLE-2016-BLG-1045][]{OB161045}. The follow-up teams included the Las Cumbres Observatory (LCO) global network, the Microlensing Follow-Up Network ($\mu$FUN, \citealt{mufun}) and Microlensing Network for the Detection of Small Terrestrial Exoplanets (MiNDSTEp, \citealt{MINDSTEp}). The LCO global network provided observations from its 1.0m telescopes located at CTIO, SAAO and SSO,  with the SDSS-$i'$ filter. The $\mu$FUN team followed the events using the 1.3m SMARTS telescope at CTIO (CT13) with V/I/H-bands \citep{CT13}, the 0.4m telescope at Auckland Observatory (AO) using a number 12 Wratten filter (which is similar to $R$-band), and the 0.36m telescope at Kumeu Observatory (Kumeu) in Auckland. The MiNDSTEp team monitored the events using the Danish 1.54-m telescope sited at ESO’s La Silla observatory in Chile, with a non-standard filter. 

%The OGLE-IV survey alerts $\sim2000$ real-time microlensing events annually by its Early Warning System \citep{Udalski1994,Udalski2003}.The MOA group alerts about 600 microlensing events per year \citep{Bond2001}.Currently, KMTNet conducts a total of (3, 7, 11, 3) fields observed at cadences $\Gamma = (4, 1, 0.4, 0.2)~ {\rm hr}^{-1}$ toward the Galactic bulge. 

%New Zealand with ?-band, the 0.36m unfiltered telescope at Possum Observatory (Pos) in New Zealand, the 0.4m telescope at Beverly-Begg Observatory (BBO) in New Zealand with $I$-band, the 0.36m unfiltered telescope at Klein Karoo Observatory (KKO) in South Africa and the 0.36m telescope at Turitea Observatory (Tur) in New Zealand using a number 12 Wratten filter.

We provide detailed descriptions of the observations for \three\ and \two\ in the next part. 

\subsection{\three}
\three\ was discovered by the OGLE collaboration on 2017 June 20. With equatorial coordinates $(\alpha, \delta)_{\rm J2000}$ = (17:41:12.65, $-26$:44:28.1) and Galactic coordinates $(\ell,b)=(1.36, 1.98)$, it lies in OGLE field BLG652, monitored by OGLE with a cadence of 0.5--1 observations per night \citep{OGLEIV}. This event was located in the gap of two CCD chips of KMTNet BLG15 field, and thus the follow-up observations were important supplements to the sparse observations from the ground-based surveys. The $I/H$-band observations from CT13 intensively covered the falling side of the peak, and its $H$-band data were also used to derive the color of the source because this event suffered from very high extinction ($A_I \sim 4.5$; See Section \ref{lens}). In addition, \three\ was also densely observed by the LCO network, the 0.4 m telescope at Auckland Observatory (AO), and the 0.36m telescope at Kumeu Observatory (Kumeu). \three\ was selected as a ``secret'' \Sp\ target on 2017 June 25 (UT 16:00) because the newest OGLE point (HJD = 2457932.78) indicated a significant rise (consistent with a high-magnification event) and the event was predict to peaked within 1 day, and it was formally announced as a \Sp\ target on 2017 June 28. The \Sp\ observations began on 2017 June 30 and ended on 2017 July 13 with 16 data points in total.
%was chosen as a \Sp\  target on 2017 June 25 (UT 16:00) because of the possibility that the event could be highly magnified. The event

\subsection{\two}
\two\ was first alerted by the OGLE collaboration on 2017 July 2. The event was located at equatorial coordinates $(\alpha, \delta)_{\rm J2000}$ = (17:57:23.56, $-27$:13:13.3), corresponding to Galactic coordinates $(\ell,b)=(2.80, $-$1.36)$. It therefore lies in OGLE field BLG645, which has a cadence less than 0.5 observations per night \citep{OGLEIV}. This event was also identified by MOA group as MOA-2017-BLG-373 $\sim$12.2 days later \citep{Bond2001}, and recognized by KMTNet's event-finding algorithm as KMT-2017-BLG-0374 \citep{Kim2018a}. The KMTNet group observed this event in its two slightly offset fields BLG02 and BLG42, with combined cadence of $\Gamma = 4~{\rm hr}^{-1}$. The LCO, $\mu$FUN, and MiNDSTEp follow-up teams also observed this event. The dense observations during the peak by LCO and MiNDSTEp were important to constrain the finite-source effects. The $H$-band observations taken by CT13 were important for characterizing the source star because this event suffered from very high extinction ($A_I \sim 4.2$; See Section \ref{lens}). \two\ was chosen as a ``secret'' \Sp\ target on 2017 July 2 (UT 20:48) because (1) the model predicted that the event could be a high-magnification event; (2) KMTNet has a high cadence of $\Gamma = 4~{\rm hr}^{-1}$. It was ``subjectively'' selected on July 6 and became "objective" on July 17 (see \citealt{YeeSpitzer}). The \Sp\ observations began on 2017 July 7 and ended on 2017 August 3 with a cadence of $\sim 1$ observation per day.  

%The event was later announced as a ``subjective" target on 2017 July 6 (UT 17:08) and became an ``objective'' target on 2017 July 17. 

%\two\ was chosen as \Sp\ a ``secret" target on 2017 July 2 (UT 20:48) based on that (1) the event could be a high-magnification event; (2) KMTNet has a high cadence of $\Gamma = 4~{\rm hr}^{-1}$. The event was later announced as a ``subjective" target on 2017 July 6 (UT 17:08) and became an ``objective'' target on 2017 July 17.  

\subsection{Data Reductions}
The photometry of OGLE, MOA, KMTNet, LCO, AO, Kumeu, and Danish data was extracted using custom implementations of the difference image analysis technique \citep{Alard1998}: \citealt{Wozniak2000} (OGLE), \citealt{Bond2001} (MOA), \citealt{pysis} (KMTNet, LCO, AO, and Kumeu), and \citealt{DanDIA} (Danish). In addition, the CT13 data were reduced by \texttt{DoPHOT}~\citep{dophot}. The \Sp\ data were reduced using the algorithm developed by \cite{Spitzerdata} for crowded-field photometry. 

%, and the Danish data were reduced via an updated version of the DanDIA pipeline \citep{DanDIA}

\section{Light curve analysis}\label{model}

\subsection{Ground-based data only}\label{ground}
For each event, we model the ground-based data using four parameters for the magnification, $A(t)$. These include three Paczy{\'n}ski parameters ($t_0, u_0, \te$) \citep{Paczynski1986} to describe the light curve produced by a single-lens with a point-source: the time of the maximum magnification as seen from Earth $t_0$, the impact parameter $u_0$ (in units of the angular Einstein radius $\thetae$), and the Einstein radius crossing time $\te$. In addition, the source size normalized by the angular Einstein radius $\rho$ is needed to incorporate finite-source effects. The flux, $f(t)$, calculated from the model is
\begin{equation}
    f(t) = f_{\rm s} A(t) + f_{\rm b},
\end{equation}
where $f_{\rm s}$ represents the flux of the source star being lensed, and $f_{\rm b}$ is any blended flux that is not lensed. The two linear parameters, $f_{\rm s}$ and $f_{\rm b}$, are different for each observatory and each filter. In addition, we adopt the linear limb-darkening law to consider the brightness profile of the source star \citep{An2002}
\begin{equation}
    S_\lambda(\theta) = \bar{S}_\lambda \left[1 - \Gamma_{\lambda}(1 - \frac{3}{2}\cos\theta)]\right],
\end{equation}
where $\bar{S}_\lambda$ is the mean surface brightness of the source, $\theta$ is the angle between the normal to the surface of the source and the line of sight, and $\Gamma_{\lambda}$ is the limb-darkening coefficient at wavelength $\lambda$.
%If it is free, we need additional one parameter for each bandpass.
%\footnote{We choose 18 as the magnitude zeropoint. Thus, the magnitude of the source can be derived by $I_{\rm s} = 18 - 2.5 * \log_{10}(f_{\rm s}$).}
We employ the Markov chain Monte Carlo (MCMC) $\chi^2$ minimization using the \texttt{emcee} ensemble sampler \citep{emcee} to find the best-fit parameters and their uncertainties.

\subsection{Satellite parallax}\label{parallax}
We measure the microlens parallax from the light curve modeling,
\begin{equation}
    \vec{\pi}_{\rm E} \sim \frac{\mathrm{AU}}{D_\perp}\left(\Delta{u_{0}}, \frac{\Delta{t_0}}{\te}\right),
\end{equation}
where $\Delta{t_0}$ is the difference in event peak time $t_0$ and $\Delta{u_{0}}$ is the difference in impact parameter $u_0$ as seen from the \Sp\ satellite and Earth, and $D_\perp$ is the projected separation between the \Sp\ satellite and Earth at the time of the event. Generally, only the absolute value of $u_0$ can be measured from the modeling, and thus the satellite parallax measurements usually suffer from a four-fold degeneracy \citep{1966MNRAS.134..315R,1994ApJ...421L..75G}. We specify the four solutions as $(+,+)$, $(+,-)$, $(-,-)$, and $(-,+)$ using the sign convention described in \cite{2015ApJ...805....8Z}. Briefly, the first and second signs in each parenthesis indicate the signs of $u_{0,\oplus}$ and $u_{0,Spitzer}$, respectively. In addition, the \Sp\ observations only cover the falling part of \three\, which leads to large uncertainty of $\pie$. Thus, we include a color-color constraint on the \Sp\ source flux $f_{s,Spitzer}$ to improve the parallax measurement \citep[e.g.][]{Novati2015}. This constraint adds a $\chi^2_{\rm penalty}$ into the total $\chi^2$ (See Equation (2) in \citealt{OB160168} for the form of the $\chi^2_{\rm penalty}$).    
%The total $\chi^2_{\rm total}$ of the model is of the form
%\begin{equation}
    %\chi^2_{\rm total} = \chi^2_{\rm }
%\end{equation}

%\subsection{Limb Darkening Effect}\label{LD}
 
%$\Gamma_{\lambda}$ values for the two events were derived based on the estimated color and magnitude of the source.

\subsection{\three}\label{model3}
Using the intrinsic color of the source star (see Section \ref{lens3}) and the color-temperature relation of \cite{Houdashelt2000}, we estimate the effective temperature of the source to be $T_{\rm eff} \approx 4450~$K. Applying ATLAS models and assuming a surface gravity of $\log g$ = 2.5, a metallicity of [M/H] = 0.0, and a microturbulence parameter of 1 ${\rm km\,s^{-1}}$, we obtain the linear limb-darkening coefficients $u_I = 0.60$ for $I$ band, $u_V = 0.81$ for $V$ band, $u_R = 0.71$ for $R$ band, $u_H = 0.39$ for $H$ band, and $u_L = 0.24$ for $L$ band \citep{Claret2011}. We then employ the transformation formula in \cite{Fields2013}, yielding the corresponding limb-darkening coefficients $\Gamma_{I} = 0.50, \Gamma_{V} = 0.74, \Gamma_{R} = 0.62, \Gamma_{H} = 0.30, {\rm and}~\Gamma_{L} = 0.18$. In the light curve modeling, we use $\Gamma_{I}$ for OGLE, LCO, CT13 $I-$band and Kumeu data, $\Gamma_{\rm AO} = (\Gamma_{V} + \Gamma_{R})/2 = 0.68$ for AO data, $\Gamma_{H}$ for CT13 $H-$band data, and $\Gamma_{L}$ for \Sp\ data.

To derive the color-color constraint on the \Sp\ source flux $f_{s,Spitzer}$, we extract the \Sp\ photometry of red giant bulge stars ($4.0 < I_{\rm OGLE} - H_{\rm VVV} < 5.5; 17.5 < I_{\rm OGLE} < 20.0$) and fit for the two parameters in the equation
\begin{equation}
    I_{\rm OGLE} - L_{\rm \Sp} = c_0 + c_1(I_{\rm OGLE} - H_{\rm VVV} - X_p),
\end{equation}
where $X_p = 4.65$ is a pivot parameter chosen to minimize the covariance between the parameters. We then obtain $c_0 = 4.47 \pm 0.01, c_1 = 1.28 \pm 0.03$. This, when combined with $(I_{\rm OGLE} - H_{\rm VVV})_{\rm S} = 4.71 \pm 0.01$ (see Section \ref{lens3}), yields $(I_{\rm OGLE} - L_{\rm \Sp})_{\rm S} = 4.55 \pm 0.02$. We employ this constraint on the light curve modeling. 
%which reduces $\sim20\%$ uncertainties on $\pi_{\rm E,N}$ and $\sim50\%$ uncertainties on $\pi_{\rm E,S}$. 

\begin{figure}[htb] 
    \includegraphics[width=\columnwidth]{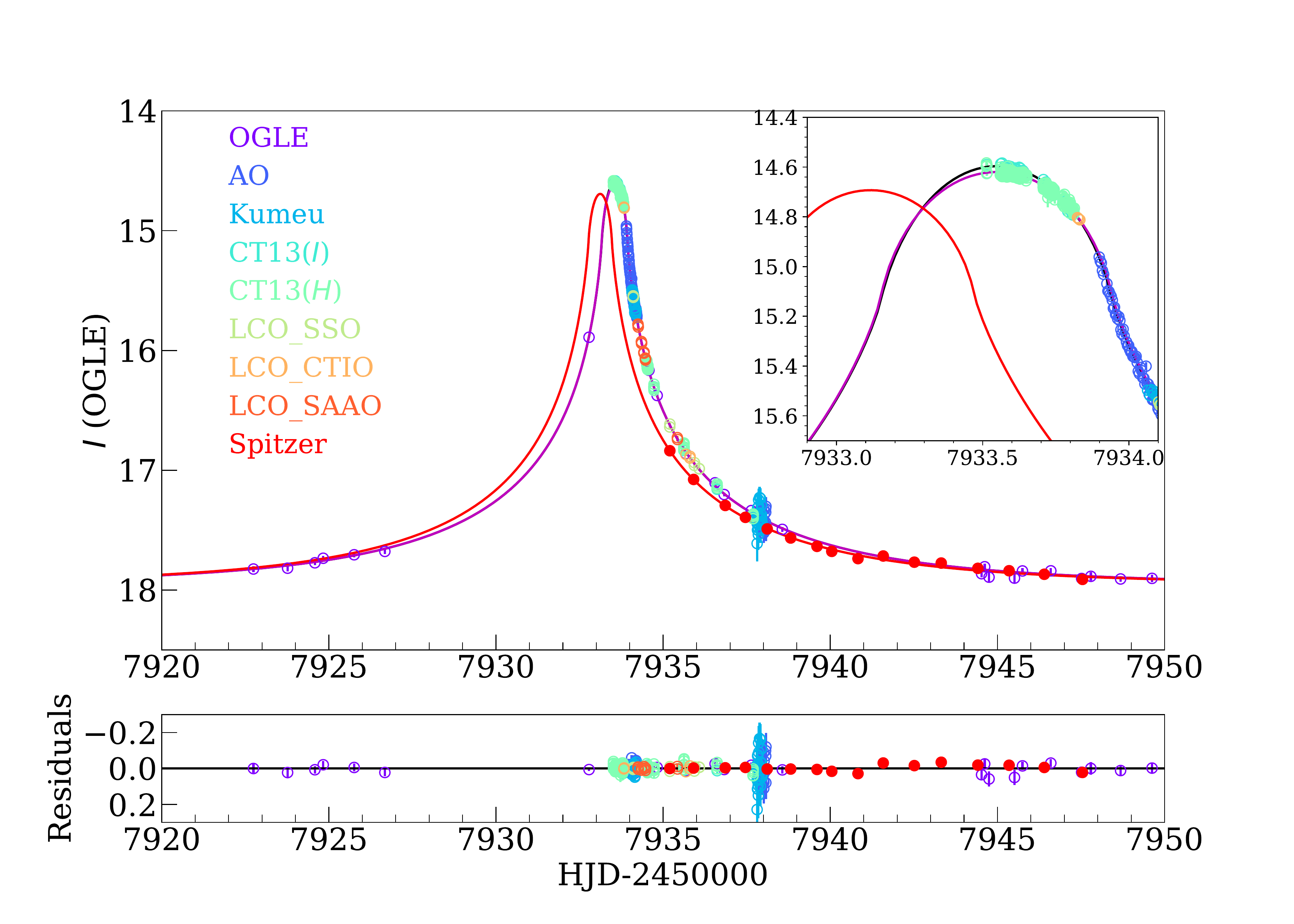}
    \caption{The light curves of event \three. The black and magenta lines represent the best-fit $(-,+)$ model for the ground data with $I$ and $H$ band, respectively, and the red line shows the corresponding model for \Sp. The inset in the top panel shows the peak of the event, with a clear finite-source effect. The circles with different colors are ground-based data points from different collaborations or bands. The red dots are \Sp\ data points.}
    \label{lc3}
\end{figure}

Table \ref{parm} shows the best-fit parameters and their $1\sigma$ uncertainties for the four-fold degenerate solutions ($\Delta\chi^2 < 0.16$). The best-fit model curves for the $(-,+)$ solution are shown in Figure \ref{lc3}. For all the four-fold degenerate solutions, the East component $\pi_{\rm E,E}$ of the microlens parallax vector is $\sim 0.038 \pm 0.06$, while the North component $\pi_{\rm E,N}$ is consistent with 0 at the $\sim 2\sigma$ level.

\subsection{\two}\label{model2}
\begin{figure}[htb] 
    \includegraphics[width=\columnwidth]{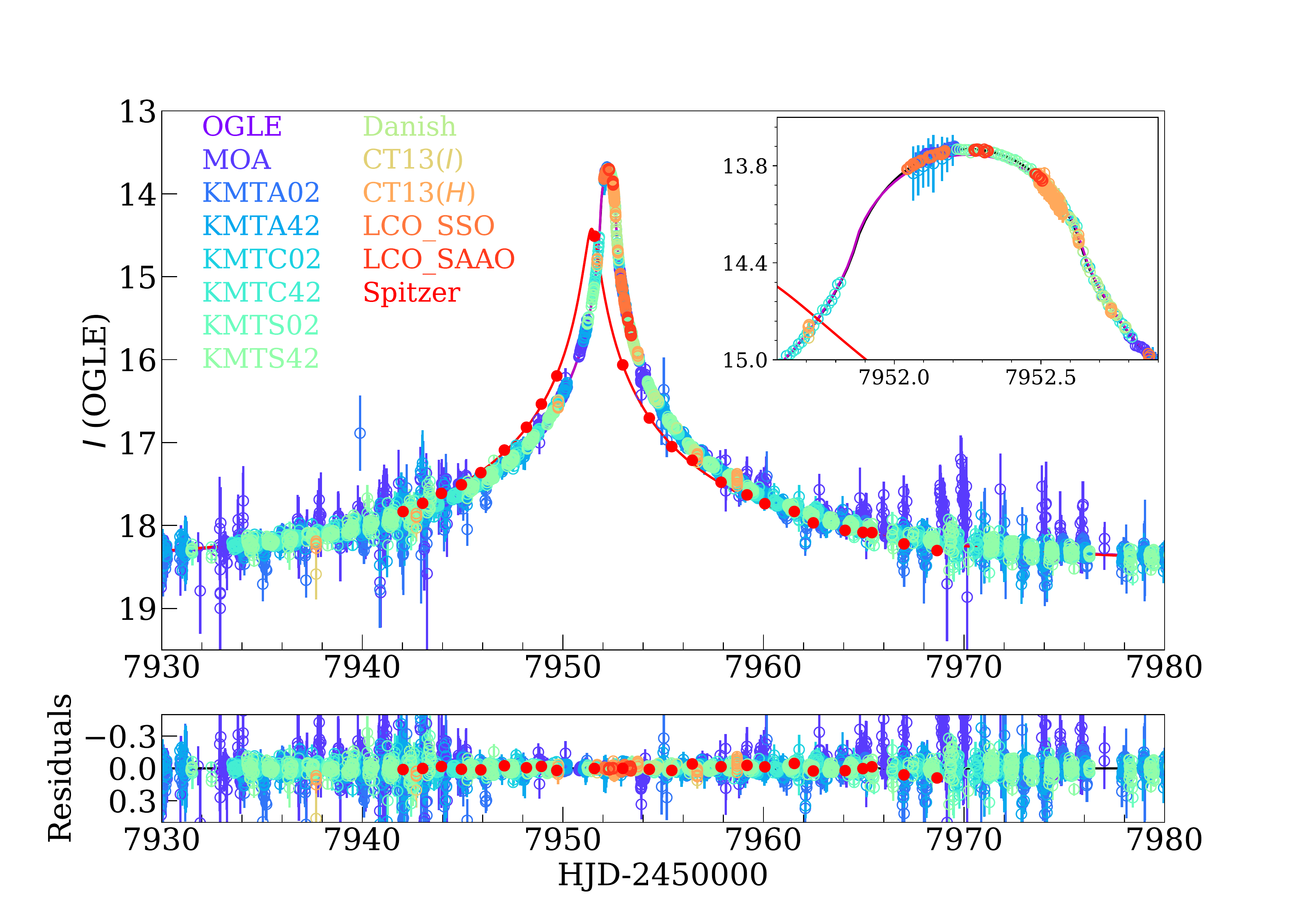}
    \caption{Ground-based and Spitzer data and best-fit model light curves of event \two\ for the $(0,+)$ model. Symbols are similar to those in Figure \ref{lc3}.}
    \label{lc2}
\end{figure}

Applying the same procedure as in Section \ref{model3}, we obtain the corresponding limb-darkening coefficients $\Gamma_{I} = 0.45, \Gamma_{R} = 0.55, \Gamma_{H} = 0.26, {\rm and}~\Gamma_{L} = 0.16$. In the light curve modeling, we use $\Gamma_{I}$ for OGLE, KMTNet, LCO, CT13 $I-$band and Danish data, $\Gamma_{\rm MOA} = (\Gamma_{I} + \Gamma_{R})/2 = 0.50$ for MOA data, $\Gamma_{H}$ for CT13 $H-$band data, and $\Gamma_{L}$ for \Sp\ data. We find that the impact parameter $u_{0,\earth} \simeq 0$, so the four degenerate solutions reduce to two solutions [$(0,+), (0,-)$], with $\Delta\chi^2 = 0.09$. However, this degeneracy has no effect on the mass and distance measurement for the lens \citep{GouldYee2012,OB161045}. The best-fit model curves for $(0,+)$ are shown in Figure \ref{lc2}, and the best-fit parameters for the two degenerate solutions are shown in Table \ref{parm}.

For this event the \Sp\ light curve precisely constrains the microlens parallax without the need of color-color constraint on $L_{\rm \Sp}$. Nevertheless, we derive the $IHL$ color-color relation using red giant stars ($3.6 < I_{\rm OGLE} - H_{\rm VVV} < 5.0; 17.0 < I_{\rm OGLE} < 19.5$) for validation of the color-color method. The relation and the $(I_{\rm OGLE} - H_{\rm VVV})_{\rm S}$ color in Section \ref{lens2} suggest $(I_{\rm OGLE} - L_{\rm \Sp})_{\rm S} = 3.82 \pm 0.03$, which is in excellent agreement with the color measured from the model of $(I_{\rm OGLE} - L_{\rm \Sp})_{\rm S} = 3.81 \pm 0.02$.

%To derive the color-color constraint on the \Sp\ source flux $f_{s,Spitzer}$, we extract \Sp\ photometry for the red giant stars ($3.6 < I - H < 4.8; 17.0 < I < 19.5$) and fit for the relation of the equation
%\begin{equation}
 %   I - L_{\rm \Sp} = c_0 + c_1(I - H - X_p),
%\end{equation}
%where $X_p = 4.30$ is a pivot parameter chosen to minimize the covariance between the parameters. We then get $c_0 = 4.06 \pm 0.01, c_1 = 1.25 \pm 0.04$. This, when combined with $(I - H)_{\rm S} = 3.97 \pm 0.03$, yields $(I - L_{\rm \Sp})_{\rm S} = 3.65 \pm 0.06$. We employ this constraint on the light curve modeling and find that $\chi^2_{\rm penalty} \sim 8.0$ $(\sim 2.8\sigma$ level) for the best-fit model. In addition, because the impact parameter as seen from Earth has $u_{0,\earth} \simeq 0$, the four degenerate solutions reduce to two solutions [(0,+), (0,-)], with $\Delta\chi^2 = 0.3$. However, this degeneracy has no effect on the mass and distance measurement for the lens \citep{GouldYee2012}. The best-fit model curves are shown in Figure \ref{lc2}, and the best-fit parameters for the two degenerate solutions are shown in Table \ref{parm2}.

% Because the source star is a bluish clump star and relatively isolated in the color-magnitude diagram (CMD), it is reasonable that the color-color relation derived via the whole clump stars not apply. Thus, we do not use the color-color constraint in the modeling.

\section{Physical parameters: Two low-mass Stars in the Galactic Bulge}\label{lens}
\subsection{{\three}L}\label{lens3}
\begin{figure*}[htbp]
    \centering
    \subfigure{
    \begin{minipage}{8.5cm}
    \centering
    \includegraphics[width=\columnwidth]{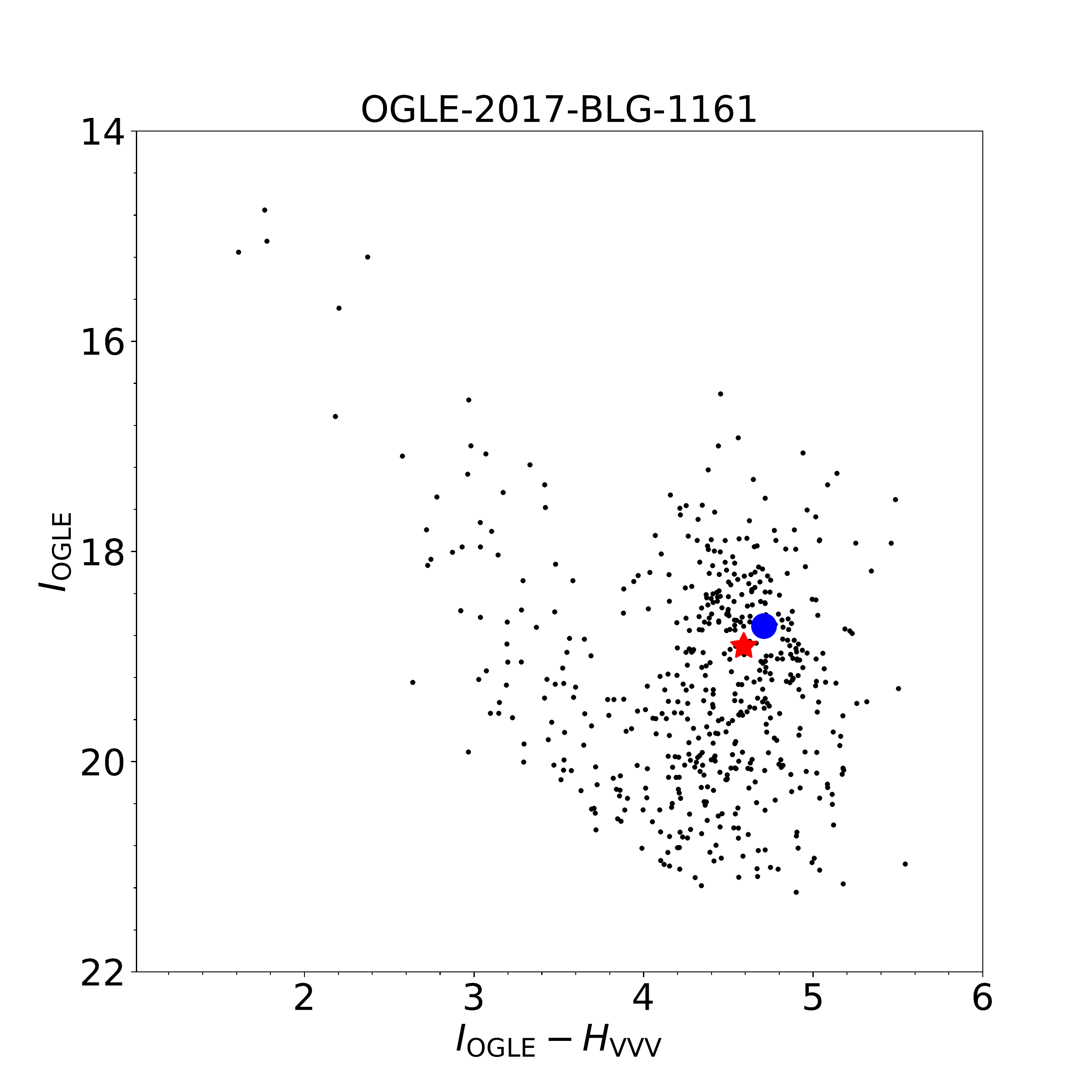}
    \end{minipage}
    }
    \subfigure{
    \begin{minipage}{8.5cm}
    \centering
    \includegraphics[width=\columnwidth]{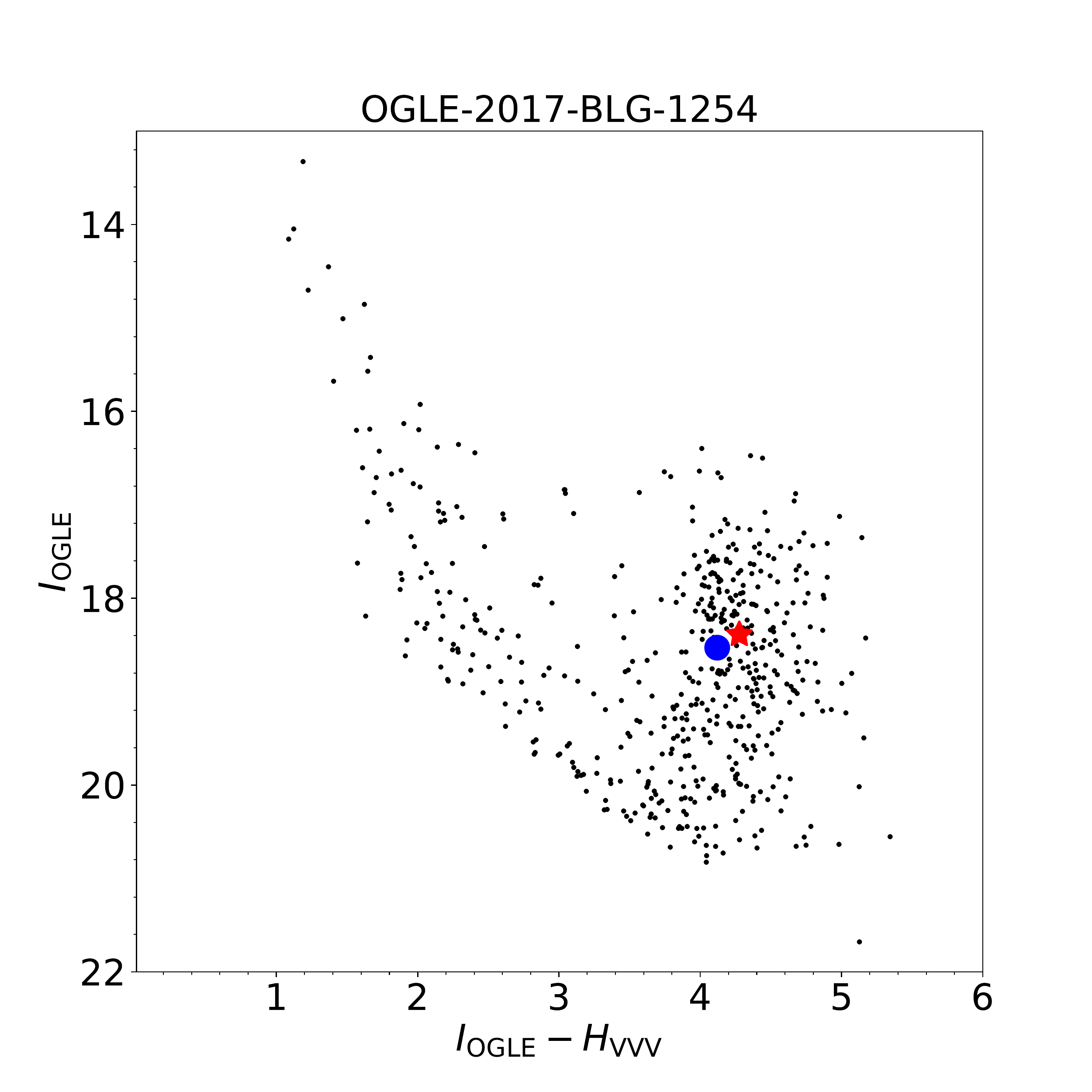}
    \end{minipage}
    }
    \caption{OGLE-VVV color-magnitude diagrams of a $2' \times 2'$ square centered on \three\ (left panel) and \two\ (right panel). The red asterisks show the centroid of the red clump. The blue dots indicate the position of the source.}
    \label{cmd}
\end{figure*}

To derive the angular Einstein radius $\thetae$ for the lens by Equation (\ref{thetae}), we estimate the angular radius $\theta_*$ of the source by locating it on a color-magnitude diagram \citep{Yoo2004}. We construct an $I - H$ versus $I$ color-magnitude diagram by cross-matching the OGLE-IV $I$-band and the VVV \citep{VVV} $H$-band stars within a $2' \times 2'$ square centered on the event (See Figure \ref{cmd}). We estimate the red giant clump to be $(I - H, I)_{\rm cl} = (4.59 \pm 0.02, 18.90 \pm 0.03)$ and find that the position of the source is $(I - H, I)_{\rm S} = (4.71 \pm 0.01, 18.70 \pm 0.03)$ from OGLE $I-$band data and CT13 $H-$band data aligned to the VVV magnitudes. From \cite{Nataf2016}, we find that the intrinsic color and de-reddened magnitude of the red clump are $(I - H, I)_{\rm cl,0} = (1.30, 14.39)$. Thus, the intrinsic color and de-reddened brightness of the source are $(I - H, I)_{\rm S,0} = (1.42 \pm 0.03, 14.19 \pm 0.04)$. These values suggest the source is a K-type giant star \citep{Bessell1998}. Using the color/surface-brightness relation of \cite{Adams2018}, we obtain 

\begin{equation}
    \theta_* = 7.4 \pm 0.4~\mu {\rm as}.
\end{equation}
We derive the angular Einstein radius and the geocentric lens-source relative proper motion
\begin{equation}
    \thetae = \frac{\theta_*}{\rho} = 0.159 \pm 0.009~\text{mas};
\end{equation}
\begin{equation}
    \mu_{\rm rel} = \frac{\thetae}{\te} = 6.11 \pm 0.39~\text{mas yr}^{-1}.
\end{equation}
Using Equation (\ref{eq:mass}), we measure the lens mass, 

\begin{numcases}{M = \frac{\thetae}{\kappa\pie} =}
0.51^{+0.12}_{-0.10} M_{\odot}~{\rm for}~\pie\simeq0.038 \\
0.38^{+0.13}_{-0.12} M_{\odot}~{\rm for}~\pie\simeq0.051.      
\end{numcases}
The lens-source relative parallax for the two cases is
\begin{numcases}{\pi_{\rm rel} =}
0.0062\pm0.0014~{\rm for}~\pie\simeq0.038 \\
0.0083\pm0.0025~{\rm for}~\pie\simeq0.051,      
\end{numcases}
which are very small compared to the source parallax $\pi_{\rm S} \simeq 0.12$ \citep{Nataf2016}. Thus, the distance between the lens and the source is determined much more precisely than the distance to the lens or the source separately. We measure the lens-source distance, 
\begin{numcases}{D_{\rm LS} \simeq D_{\rm S}^{2}\frac{\pi_{\rm rel}}{\rm AU}}
0.40\pm0.12~{\rm kpc~for}~\pie\simeq0.038 \\
0.53\pm0.19~{\rm kpc~for}~\pie\simeq0.051,      
\end{numcases}
where we adopt the source distance $D_{\rm S} = 8.0 \pm 0.8$~kpc using the Galactic model of \cite{Zhu2017spitzer}. Because the lens-source distance is $\lesssim 1$~kpc and the source is almost certainly a bulge red-clump star, the lens should be an M/K dwarf in the Galactic bulge. We list the derived source star properties in Table \ref{source} and the physical parameters of all the four-fold degenerate solutions in Table \ref{phy}. 
%\begin{equation}
    %M_{\rm L} = \frac{\thetae}{{\kappa}\pie} = 0.34^{+0.30}_{-0.11} M_{\odot};
%\end{equation}
%

%\begin{equation}
    %D_{\rm L} = \frac{\mathrm{au}}{\pie\thetae + \pi_{\rm S}} = 7.52^{+0.21}_{-0.20}~\text{kpc},
%\end{equation}

%The lens mass and distance are 
%\begin{equation}
 %   M_{\rm L} = 0.30 \pm 0.12~M_{\odot}, \,D_{\rm L} = 7.44 \pm 0.16~{\rm kpc}    
%\end{equation}
%or 
%\begin{equation}
 %   M_{\rm L} = 0.17 \pm 0.05~M_{\odot}, \,D_{\rm L} = 7.03 \pm 0.22~{\rm kpc}.    
%\end{equation}

%We also construct an $I - H$ versus $I$ color-magnitude diagram via the OGLE-IV $I-$band and the VVV  $H-$band stars within $2'$ separation from the event (See Figure \ref{cmd3}). We measure the centroid of the red giant clump $(I - H, I)_{\rm cl} = (4.59 \pm 0.02, 18.90 \pm 0.03)$ and the position of the source $(I - H, I)_{\rm S} = (4.63 \pm 0.03, 18.69 \pm 0.02)$. From \cite{Nataf2016}, we find that the intrinsic color and de-reddened magnitude of the red clump are $(I - H, I)_{\rm cl,0} = (1.32, 14.39)$, from which we derive the intrinsic color and de-reddened brightness of the source are $(I - H, I)_{\rm S,0} = (1.36 \pm 0.04, 14.18 \pm 0.04)$. Thus, the source is a typical red giant star. Applying the color/surface-brightness relation in \cite{Adams2018}, we finally obtain 

\subsection{{\two}L}\label{lens2}
We construct an $I - H$ versus $I$ color-magnitude diagram via the OGLE-IV $I$-band and the VVV $H$-band stars within a $2' \times 2'$ square centered on the event (See Figure \ref{cmd}). We measure the centroid of the red giant clump $(I - H, I)_{\rm cl} = (4.28 \pm 0.02, 18.39 \pm 0.03)$ and the position of the source $(I - H, I)_{\rm S} = (4.12 \pm 0.02, 18.53 \pm 0.01)$. From \cite{Nataf2016}, we find that the intrinsic color and de-reddened magnitude of the red clump are $(I - H, I)_{\rm cl,0} = (1.30, 14.35)$, from which we derive the intrinsic color and de-reddened brightness of the source are $(I - H, I)_{\rm S,0} = (1.14 \pm 0.03, 14.51 \pm 0.03)$. Thus, the source is a G-type giant star \citep{Bessell1998}. Applying the color/surface-brightness relation of \cite{Adams2018}, we obtain

\begin{equation}
    \theta_* = 5.2 \pm 0.3~\mu {\rm as};
\end{equation}

\begin{equation}
    M_{\rm L} = \frac{\thetae}{{\kappa}\pie} = 0.60 \pm 0.03 M_{\odot};
\end{equation}
\begin{equation}
     D_{\rm LS} \simeq D_{\rm S}^{2}\frac{\pi_{\rm rel}}{\rm AU} = 0.53 \pm 0.11~\text{kpc},
\end{equation}
where we also adopt the source distance $D_{\rm S} = 7.8 \pm 0.8$~kpc using the Galactic model of \cite{Zhu2017spitzer}. Thus, the lens is probably a K dwarf in the Galactic bulge. We list the derived source star properties in Table \ref{source} and the physical parameters of \two\ in Table \ref{phy}.

%From the light curve modeling, we obtain $\rho = 0.0250 \pm 0.0006$. This indicates the angular Einstein radius and the lens-source relative proper motion
%\begin{equation}
 %   \thetae = \frac{\theta_*}{\rho} = 0.180 \pm 0.009~\text{mas};
%\end{equation}
%\begin{equation}
 %   \mu_{\rm rel} = \frac{\thetae}{\te} = 4.26 \pm 0.230~\text{mas yr}^{-1}.
%\end{equation}

%Combining this with the microlens parallax parameter $\pie = 0.042 \pm 0.001$, we obtain 

%To derive the angular Einstein radius $\thetae$ for the lens by Equation \ref{thetae}, we estimate the angular radius $\theta_*$ of the source by placing the source on a color-magnitude diagram (CMD) \cite{Yoo2004}. We construct an $I - H$ versus $I$ color-magnitude diagram by cross-matching the OGLE-IV $I-$band and the VVV \citep{VVV} $H-$band stars within $2'$ square centered on the event (See Figure \ref{cmd2}). We estimate the red giant clump to be $(I - H, I)_{\rm cl} = (4.30 \pm 0.02, 18.38 \pm 0.03)$ and find the position of the source is $(I - H, I)_{\rm S} = (3.97 \pm 0.03, 18.54 \pm 0.02)$ from OGLE $I-$band data and CT13 $H-$band data. From \cite{Nataf2016}, we find that the intrinsic color and de-reddened magnitude of the red clump are $(I - H, I)_{\rm cl,0} = (1.32, 14.35)$. Thus, the intrinsic color and de-reddened brightness of the source are $(I - H, I)_{\rm S,0} = (0.99 \pm 0.04, 14.51 \pm 0.04)$. These values suggest the source star is a bluish clump star. Using the color/surface-brightness relation in \cite{Adams2018}, we finally find 

\section{Discussion and Conclusion}\label{dis}
We have reported the analysis of two microlensing events \three\ and \two, each of which displays both finite-source effects detected by the ground-based data and the microlens parallax measured by the joint analysis of the ground-based data and the \Sp\ data. Including these two events, the \Sp\ microlensing program has measured the mass and distance for eight isolated objects from 2015--2017, yielding an estimate of the apparent detection frequency $\sim 8/328 = 2.4\%$\footnote{\Sp\ observed 524 events from 2015--2017, but only 328 events are single-lens events with a clear \Sp\ signal.}. This apparent frequency agrees with the theoretical frequency $\sim 3.3\%$ \citep{ZhuPLFS} within $1\sigma$ for Poisson statistics. The theoretical frequency assumes that the probability to detect the finite-source effects in single-lens events is the same for ground and \Sp\ observations, but the \Sp\ data only detected finite-source effects for two events\footnote{For OGLE-2017-BLG-1186, the best-fit \Sp\ light curve also shows finite-source effects, but the daily \Sp\ data are insufficient for the detection.} (OGLE-2015-BLG-0763 \citep{ZhuPLFS}, OGLE-2015-BLG-1482 \citealt{OB151482}), with a degeneracy in $\rho$. This is because the \Sp\ observations only have a $\Gamma \sim {\rm day}^{-1}$ cadence and require a 3--10 day turnaround time after selection of the event, leading to the loss of finite-source effect detection from \Sp\ observations. 

%suggesting a possible 2 factor deficit., which might be due to the incomplete coverage of the \Sp\ observations in some events.

The probability of finite-source effects occurring in a single-lens event is 
\begin{equation}
    P = \rho \equiv \frac{\theta_*}{\thetae}. %\propto \theta_*.
\end{equation}
This, when combined with the microlensing rate $\Gamma_{\rm \mu lens} \propto n \mu_{\rm rel}\thetae$ ($n$ is the number density), yields the finite-source event rate \citep{GouldYee2012, OB170896}
\begin{equation}
    \Gamma_{\rm FS} = \rho \Gamma_{\rm \mu lens} \propto n \mu_{\rm rel} \theta_*.
\label{FS}
\end{equation}

\begin{figure}[htb] 
    \includegraphics[width=\columnwidth]{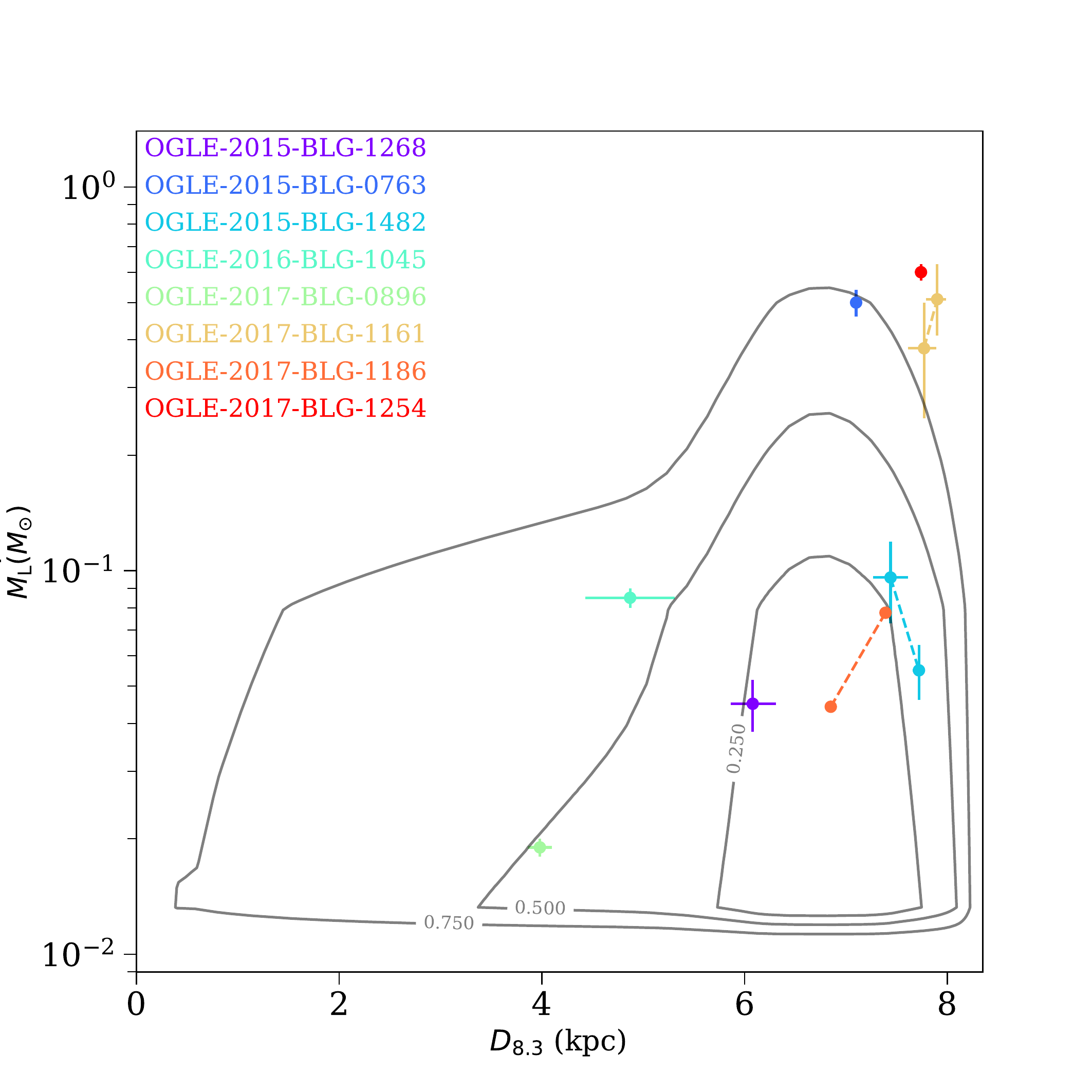}
    \caption{Bayesian probability density distributions from the Galactic model of \cite{Zhu2017spitzer} compared to the eight published \Sp\ finite-source events. We fix the source distance to 8.3~kpc and then derive the lens distance $D_{8.3}$ for all the events. The predicted mass distribution is derived from the initial mass function of \cite{Kroupa2001}. The dots with different colors represent different events. The two dots connected by dash lines represent the two degenerate solutions of one event. The grey lines represent equal probability density. The values on the contours indicate the total probability inside the contours predicted by the Galactic model and the total probability is normalized to unity.}
    \label{sta}
\end{figure}

\begin{figure}[htb] 
    \includegraphics[width=\columnwidth]{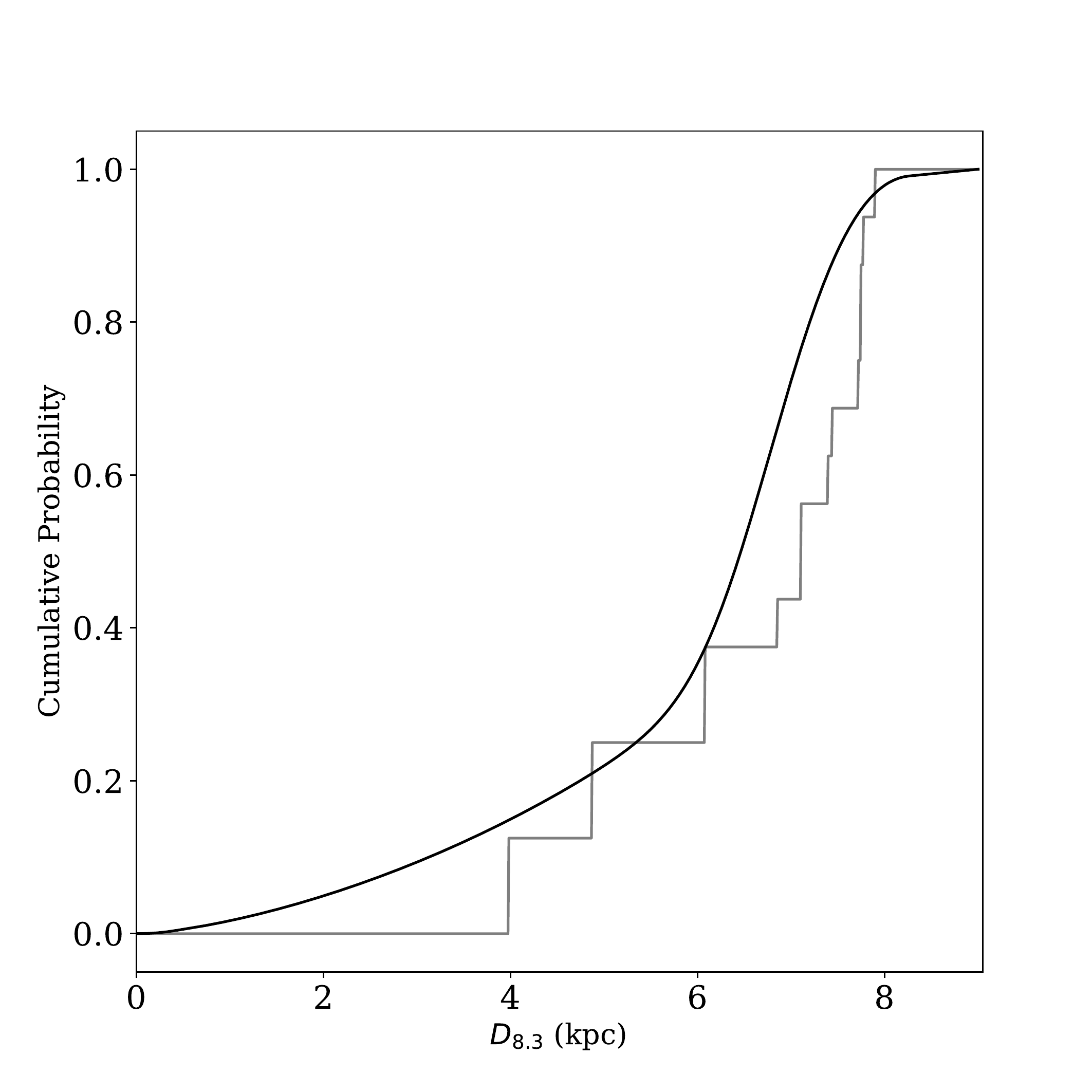}
    \caption{Cumulative distribution of the lens distance from the Galactic model of \cite{Zhu2017spitzer} and the eight published \Sp\ finite-source events. We fix the source distance of 8.3~kpc and then derive the lens distance $D_{8.3}$ for all the events. The black line represents the distribution predicted by the Galactic model, and the grey lines represents the distribution calculated from the eight events. The observed distribution is consistent with the Galactic model with a Kolmogorov-Smirnov probability of 30.3\%.}
    \label{distance}
\end{figure}

\begin{figure}[htb] 
    \includegraphics[width=\columnwidth]{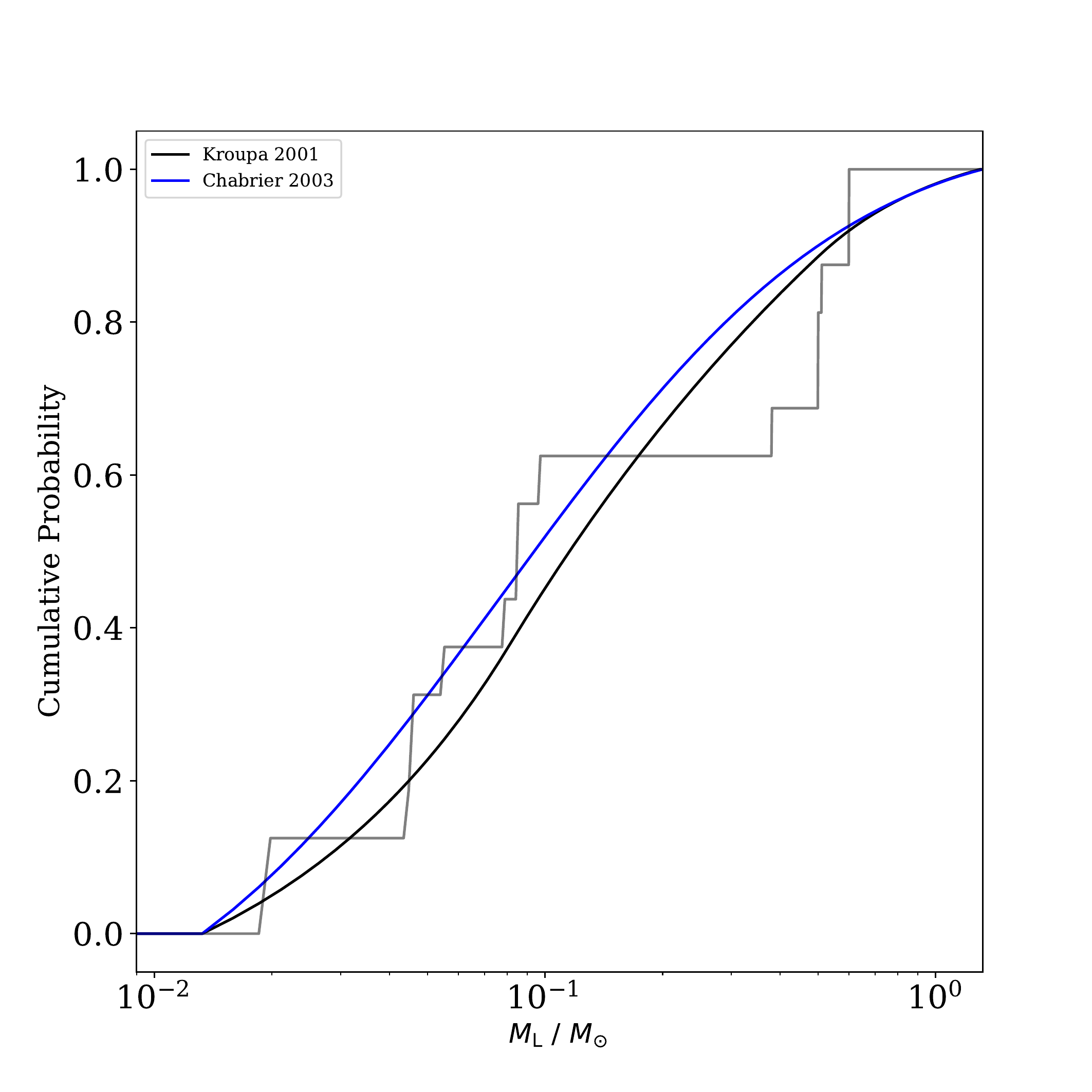}
    \caption{Cumulative distribution of the lens mass from the initial mass function and the eight published \Sp\ finite-source events. The black line represents the distribution predicted by the initial mass function of \cite{Kroupa2001} and the blue line represents the distribution calculated from \cite{Charbrier2003}. The observed distribution is consistent with the initial mass functions of \cite{Kroupa2001} and \cite{Charbrier2003} with a Kolmogorov-Smirnov probability of 84.9\% and 72.3\%, respectively.}
    \label{mass}
\end{figure}

We apply the Galactic model described in \cite{Zhu2017spitzer} and estimate the probability density distribution of finite-source events based on $n \times \mu_{\rm rel}$. We average the distributions in the direction of the eight \Sp\ finite-source events and assume the source distances are 8.3~kpc for all the events (following \citealt{Zhu2017spitzer}). For events with two degenerate solutions, both solutions are included at half the weight. Figure \ref{sta} compares the resulting probability densities for different masses and distances with the eight \Sp\ finite-source events. Figure \ref{distance} and \ref{mass} compare the cumulative distributions of the lens distance and lens mass, respectively. In this comparison, we do not take into account the \Sp\ detection efficiency, and possible selection or publication biases. Such detailed analysis is beyond the scope of this paper and will be done in future complete statistical analysis of the \Sp\ campaigns. 

The observed \Sp\ sample agrees with expectations from the Galactic model. The distance distribution of the eight events is consistent with the Galactic model of \cite{Zhu2017spitzer} with a Kolmogorov-Smirnov probability of 30.3\%, and the mass distribution is consistent with the initial mass function of \cite{Kroupa2001} and \cite{Charbrier2003} with a Kolmogorov-Smirnov probability of 84.9\% and 72.3\%, respectively. Both the Galactic model and the eight \Sp\ events show that the finite-source effects have strong bias toward objects in the Galactic bulge. This is primarily because the stellar number density in the Galactic bulge is significantly higher than that of the Galactic disk, while the lens-source relative proper motions of disk lenses are only slightly higher on average (see Figure 1 and 2 of \citealp{Zhu2017spitzer}). In addition, the finite-source effects are biased toward the more common low-mass objects (M-dwarfs and brown dwarfs). However, \Sp\ has no detection of a low-mass brown dwarf ($M_{\rm L} < 0.04 M_{\odot}$) in the Galactic bulge, in tension with the expectations from the Galactic model. This is likely due to the 3--10 day delay of the \Sp\ observations, which is comparable to the typical microlens timescale for a bulge low-mass brown dwarf is less than 6 days. 

\cite{OB140962} compared 13 well-characterized \Sp\ systems (10 binary/planetary lenses and 3 single lenses) with Bayesian predictions from Galactic models and found that they are in excellent agreement. Our preliminary comparison of eight \Sp\ single lenses also suggests good agreement with the expectations from the Galactic model. Assuming the empirical rate from 2015--2017 season, we expect another 5--10 detections of finite-source events in 2018 and 2019 \Sp\ microlensing campaigns, and thus future statistical analyses of all \Sp\ finite-source events will potentially allow a study of specific stellar populations and test the Galactic model.

\acknowledgments
W.Z., W.T., S.-S.L. and S.M. acknowledges support by the National Science Foundation of China (Grant No. 11821303 and 11761131004). This work is based (in part) on observations made with the \Sp\ Space Telescope, which is operated by the Jet Propulsion Laboratory, California Institute of Technology under a contract with NASA. Support for this work was provided by NASA through an award issued by JPL/Caltech. The OGLE has received funding from the National Science Centre, Poland, grant MAESTRO 2014/14/A/ST9/00121 to AU. This research has made use of the KMTNet system operated by the Korea Astronomy and Space Science Institute (KASI) and the data were obtained at three host sites of CTIO in Chile, SAAO in South Africa, and SSO in Australia. The MOA project is supported by JSPS KAKENHI Grant Number JSPS24253004, JSPS26247023, JSPS23340064, JSPS15H00781, JP16H06287, and JP17H02871. The research has made use of data obtained at the Danish 1.54m telescope at ESO’s La Silla Observatory. CITEUC is funded by National Funds through FCT - Foundation for Science and Technology (project: UID/Multi/00611/2013) and FEDER - European Regional Development Fund through COMPETE 2020 - Operational Programme Competitiveness and Internationalization (project: POCI-01-0145-FEDER-006922). Work by AG was supported by AST-1516842 from the US NSF and JPL grant 1500811. Wei Zhu was supported bythe Beatrice and Vincent Tremaine Fellowship at CITA. Work by CH was supported by the grant (2017R1A4A1015178) of National Research Foundation of Korea. YT acknowledges the support of DFG priority program SPP 1992 “Exploring the Diversity of Extrasolar Planets” (WA 1047/11-1). L.M. acknowledges support from the Italian Minister of Instruction, University and Research (MIUR) through FFABR 2017 fund. 

%(OGLE-2015-BLG-1268 and OGLE-2015-BLG-0763 \citep{ZhuPLFS}, OGLE-2015-BLG-1482 \citep{OB151482}, OGLE-2016-BLG-1045 \citep{OB161045}, OGLE-2017-BLG-0896 \citep{OB170896}, \three\ and \two\ (in this work), OGLE-2017-BLG-1186).

\bibliography{Zang.bib}

%\begin{table}[htb]
 %   \centering
 %   \caption{Best-fit parameters for \three}
 %   \begin{tabular}{c|c|c|c|c|c|c|c|c|c|c|c}
 %   \hline
 %   \hline
 %   Solutions & $\chi^2$/dof & $\chi^2_{\rm penalty}$ & $t_{0,\earth}$ -2450000(d) & $u_{0,\earth}$ & $\te$ & $\rho(10^{-2})$ & $\pi_{\rm E, N}$ & $\pi_{\rm E, E}$ & $\pi_{\rm E}$ & $I_{\rm s, OGLE}$ & $I_{\rm b, OGLE}$ \\
 %   \hline
 %   \hline
 %   $(+,+)$ & 618.41/617 & 0.00 & 7933.548(2) & 0.0214(8) & 9.5(3) & 4.64(15) & -0.000(23) & 0.039(7) & 0.039(9) & 18.71(3) & 18.71(3) \\
 %   $(+,-)$ & 618.35/617 & 0.02 & 7933.548(2) & 0.0214(9) & 9.5(3) & 4.65(15) & -0.034(21) & 0.038(5) & 0.051(17) & 18.70(3) & 18.72(3) \\
 %   $(-,-)$ & 618.25/617 & 0.01 & 7933.549(2) & -0.0214(9) & 9.5(3) & 4.67(15) & -0.000(22) & 0.038(5) & 0.038(8) & 18.70(3) & 18.72(3) \\
 %   $(-,+)$ & 618.30/617 & 0.02 & 7933.548(2) & -0.0214(9) & 9.4(3) & 4.66(16) & 0.037(22) & 0.037(6) & 0.052(16) & 18.70(3) & 18.72(3) \\
 %   \hline
 %   \hline
 %    $(0,+)$ & 8254.78/8256 & - & 7952.2519(4) & 0.0003(10) & 15.43(6) & 2.51(1) & 0.0203(7) & 0.0368(4) & 0.0420(7) & 18.53(1) & 21.32(6) \\
 %   $(0,-)$ & 8254.69/8256 & - & 7952.2518(4) & -0.0003(9) & 15.42(7) & 2.51(1) & -0.0174(7) & 0.0384(5) & 0.0421(7) & 18.53(1) & 21.32(6) \\ 
 %   \hline
 %   \hline
 %   \end{tabular}\\
 %   \label{parm3}
%\end{table}
%\multicolumn{12}{c}{OGLE-2017-BLG-1161}\\
\begin{table}[htb]
    \centering
    \caption{Best-fit parameters for \three\ and \two\ and their $68\%$ uncertainty range from the MCMC}
    \begin{tabular}{c|c c c c|c c}
    \hline
    \hline
     Event & \multicolumn{4}{|c|}{OGLE-2017-BLG-1161} & \multicolumn{2}{c}{OGLE-2017-BLG-1254} \\
    \hline
    Solution  & $(+,+)$ & $(+,-)$ & $(-,-)$ & $(-,+)$ & $(0,+)$ & $(0,-)$ \\
    \hline
    $t_{0,\earth}$ -2450000(d) & 7933.548(2) & 7933.548(2) & 7933.548(2) & 7933.548(2) & 7952.2519(4) & 7952.2518(4) \\
    $u_{0,\earth}$ & 0.0214(8) & 0.0214(9) & -0.0214(9) & -0.0214(9) & 0.0003(10) & -0.0003(9) \\
    $\te$ & 9.5(3) & 9.5(3) & 9.5(3) & 9.4(3) & 15.43(6) & 15.42(7) \\
    $\rho)$ & 0.0464(15) & 0.0465(15) & 0.0467(15) & 0.0466(16) & 0.0251(1) & 0.0251(1) \\
    $\pi_{\rm E, N}$ & -0.000(23) & -0.034(21) & -0.000(22) & 0.037(22) & 0.0203(7) & -0.0174(7) \\
    $\pi_{\rm E, E}$ & 0.039(7) & 0.038(5) & 0.038(5) & 0.037(6) & 0.0368(4) & 0.0384(5) \\
    $\pi_{\rm E}$ & 0.039(9) & 0.051(17) & 0.038(8) & 0.052(16) & 0.0420(7) & 0.0421(7) \\
    $I_{\rm s, OGLE}$ & 18.71(3) & 18.70(3) & 18.70(3) & 18.70(3) & 18.53(1) & 18.53(1) \\
    $I_{\rm b, OGLE}$ & 18.71(3) & 18.72(3) & 18.72(3) & 18.72(3) & 21.32(6) & 21.32(6) \\
    $\chi^2_{\rm penalty}$ & 0.00 & 0.02 & 0.01 & 0.02 & - & - \\
    $\chi^2$/dof & 618.41/617 & 618.35/617 & 618.25/617 & 618.30/617 & 8254.78/8256 & 8254.69/8256 \\
    \hline
    \hline
    \end{tabular}\\
    \label{parm}
\end{table}

\begin{table}[htb]
    \centering
    \caption{Derived Source Star Properties for \three\ and \two.}
    \begin{tabular}{c c c c}
    \hline
    \hline
    Parameters & Units & \multicolumn{2}{c}{Value}\\
      & & \three\ & \two\ \\
    \hline
    $A_I$ & [mag] & $\sim 4.5$  & $\sim 4.2$ \\
    $I_{\rm S}$ & [mag] & $18.70 \pm 0.03$ & $18.53 \pm 0.01$ \\
    $H_{\rm S}$ & [mag] & $13.99 \pm 0.03$ & $14.41 \pm 0.01$ \\
    $(I - H)_{\rm S}$ &  & $4.71 \pm 0.01$ & $4.12 \pm 0.02$ \\
    $(I - L)_{\rm S}$ &  & $4.55 \pm 0.02$ & $3.81 \pm 0.02$ \\
    $I_{\rm S,0}$ & [mag] & $14.19 \pm 0.04$ & $14.51 \pm 0.03$ \\
    $H_{\rm S,0}$ & [mag] & $12.77 \pm 0.04$ & $13.37 \pm 0.03$ \\
    $(I - H)_{\rm S,0}$ &  & $1.42 \pm 0.03$ & $1.14 \pm 0.03$ \\
    $\theta_*$ & [$\mu$as] & $7.4 \pm 0.04$ & $5.2 \pm 0.03$ \\
    \hline
    \hline
    \end{tabular}\\
    \label{source}
\end{table}

\begin{table}[htb]
    \centering
    \caption{Physical parameters for \three\ and \two.}
    \begin{tabular}{c|c c c c|c}
    \hline
    \hline
     Event & \multicolumn{4}{|c|}{\three} & \two \\
    \hline
    Solution  & $(+,+)$ & $(+,-)$ & $(-,-)$ & $(-,+)$ & $(0,+)$ and $(0,-)$ \\
    \hline
    $\thetae$ [mas] & $0.159\pm0.009$ & $0.159\pm0.009$ & $0.159\pm0.009$ & $0.159\pm0.009$ & $0.207\pm0.008$ \\
    $M_{\rm L}$ [$M_{\odot}$] & $0.50^{+0.12}_{-0.10}$ & $0.38^{+0.13}_{-0.12}$ & $0.51^{+0.11}_{-0.10}$ & $0.38^{+0.12}_{-0.11}$ & $0.60\pm0.03$ \\
    $D_{\rm LS}$ [kpc] & $0.40\pm0.12$ & $0.53\pm0.19$ & $0.40\pm0.12$ & $0.53\pm0.19$ & $0.53 \pm 0.11$ \\
    $\mu_{\rm rel}$ [$\rm mas\,yr^{-1}$] & $6.11\pm0.39$ & $6.11\pm0.39$ & $6.11\pm0.39$ & $6.11\pm0.39$ & $4.90\pm0.20$\\
    \hline
    \hline
    \end{tabular}\\
    \label{phy}
\end{table}

%\begin{table}[htb]
 %   \centering
 %   \caption{Physical parameters for \three}
 %   \begin{tabular}{c|c|c|c|c}
 %   \hline
 %   \hline
 %   Solutions & $\thetae$ [mas] & $M_{\rm L}$ [$M_{\odot}$] & $D_{\rm L}$ [kpc] ($D_{\rm S}=8.0$~kpc) & $\mu_{\rm rel}$ [$\rm mas\,yr^{-1}$]\\
 %   \hline
 %   $(+,+)$ & $0.159\pm0.009$ & $0.50^{+0.12}_{-0.10}$ & $7.62^{+0.08}_{-0.18}$ & $6.11\pm0.39$ \\
 %   $(+,-)$ & $0.159\pm0.009$ & $0.38^{+0.13}_{-0.12}$ & $7.51^{+0.12}_{-0.22}$ & $6.11\pm0.39$ \\
 %   $(-,-)$ & $0.159\pm0.009$ & $0.51^{+0.11}_{-0.10}$ & $7.63^{+0.07}_{-0.17}$ & $6.11\pm0.39$ \\
 %   $(-,+)$ & $0.159\pm0.009$ & $0.38^{+0.12}_{-0.11}$ & $7.51^{+0.12}_{-0.21}$ & $6.11\pm0.39$ \\
 %   \hline
 %   \hline
 %   \end{tabular}\\
 %   \label{phy3}
%\end{table}

%\input{table2.tex}

%\begin{figure*}[htbp]
 %   \centering
  %  \subfigure{
  %  \begin{minipage}{8.5cm}
  %  \centering
  % \includegraphics[width=\columnwidth]{OB171161.pdf}
  %  \end{minipage}
  %  }
  %  \subfigure{
  %  \begin{minipage}{8.5cm}
  %  \centering
  %  \includegraphics[width=\columnwidth]{OB171254.pdf}
  %  \end{minipage}
  %  }
  %  \caption{Bayesian posterior probability density distributions (PDFs) of the finite-source events rate for different lens distances. The left panel and the right panel are the distributions of the directions toward \three\ and \two, respectively.}
 %   \label{Bayesian}
%\end{figure*}

%\begin{figure}[htb] 
 %   \includegraphics[width=\columnwidth]{sta-2.pdf}
  %  \caption{Bayesian probability density distributions from the Galactic model of \cite{Zhu2017spitzer} compared to the eight published \Sp\ finite-source events. %The four blue dots represent events OGLE-2015-BLG-0763, OGLE-2015-BLG-1268, OGLE-2016-BLG-1045 and OGLE-2017-BLG-0896. The two magenta dots represent the two degenerate solutions of OGLE-2015-BLG-1482, and two cyan dots represent the two degenerate solutions of OGLE-2017-BLG-1186. The two green dots represent the two degenerate solutions of OGLE-2017-BLG-1161 and the red dot represent the event OGLE-2017-BLG-1254 from this work. We fix the source distance on 8.3~kpc and then derive the lens distance $D_{8.3}$ for all the events. The total probability is normalized to one.}
  %  \label{sta}
%\end{figure}

%\begin{figure}[htb] 
 %   \includegraphics[width=\columnwidth]{tE_com.pdf}
 %   \caption{Microlens timescale $\te$ distribution of \Sp\ and all events in the OGLE Early Warning System during 2015-2017. The histogram with magenta color %represents the distribution obtained from \Sp\ events, while the histogram without color is from all the OGLE events. We directly adopt the fitting parameters shown in the OGLE Early Warning System. Events with $u_{0}<0.01$ and $\te < 2$~days have been excluded because of their unreliable parameters \citep{mufun}.}
  %  \label{tE}
%\end{figure}

\end{document}